\documentclass[conference]{IEEEtran}
\IEEEoverridecommandlockouts
\usepackage[utf8]{inputenc} 
\usepackage[T1]{fontenc}    
\usepackage{hyperref}       
\usepackage{url}            
\usepackage{booktabs}       
\usepackage{amsfonts}       
\usepackage{nicefrac}       
\usepackage{microtype}      
\usepackage{xcolor}         
\usepackage[ruled,vlined]{algorithm2e}
\usepackage[nointegrals]{wasysym}

\usepackage{amsmath}
\usepackage[shortlabels]{enumitem}
\usepackage{amssymb}
\usepackage{amsmath}
\usepackage{amsthm}
\usepackage{graphicx} 
\usepackage{epstopdf} 
\usepackage{wrapfig}
\usepackage{bm}
\usepackage{caption}

\usepackage{cite}
\usepackage{amsmath,amssymb,amsfonts}
\usepackage{algorithmic}
\usepackage{graphicx}
\usepackage{textcomp}
\usepackage{xcolor}
\def\BibTeX{{\rm B\kern-.05em{\sc i\kern-.025em b}\kern-.08em
    T\kern-.1667em\lower.7ex\hbox{E}\kern-.125emX}}
\begin{document}

\title{Tolerating Adversarial Attacks and Byzantine Faults in Distributed Machine Learning
}

\author{\IEEEauthorblockN{Yusen Wu$^1$, Hao Chen$^1$, Xin Wang$^1$, Chao Liu$^1$, Phuong Nguyen$^{1,2}$, Yelena Yesha$^{1,3}$}
\IEEEauthorblockA{\textit{$^1$University of Maryland, Baltimore County, Baltimore, MD, USA} \\
\textit{$^2$OpenKneck Inc, Halethorpe, MD, USA} \\
\textit{$^3$University of Miami, FL, USA}\\ 
\{ywu5, chenhao1, xinwang11, chaoliu717, phuong3, yeyesha\}@umbc.edu}
}

\maketitle

\begin{abstract}
Adversarial attacks attempt to disrupt the training, retraining, and utilizing of artificial intelligent and machine learning models in large-scale distributed machine learning systems. This causes security risks on its prediction outcome. For example, attackers attempt to poison the model by either presenting inaccurate misrepresentative data or altering the models’ parameters. In addition, Byzantine faults including software, hardware, network issues occur in distributed systems which also lead to a negative impact on the prediction outcome. In this paper, we propose a novel distributed training algorithm, partial synchronous stochastic gradient descent (ParSGD), which defends adversarial attacks and/or tolerates Byzantine faults. We demonstrate the effectiveness of our algorithm under three common adversarial attacks again the ML models and a Byzantine fault during the training phase. Our results show that using ParSGD, ML models can still produce accurate predictions as if it is not being attacked nor having failures at all when almost half of the nodes are being compromised or failed. We will report the experimental evaluations of ParSGD in comparison with other algorithms.  
\end{abstract}


\begin{table*}[th]
\centering
\footnotesize
\begin{tabular}{|c|c|c|c|c|}
\hline
\textbf{Byzantine-resilient GARs} & \!\!\textbf{Fault Tolerance}\!\! & \!\!\textbf{Mode}\!\! &\!\!\textbf{Assumption}\!\!& \!\!\textbf{Time Complexity}\!\!\\
\hline
\hline
Median and Trimmed-mean~\cite{yin2018byzantine} & $n >= 2f + 2$ & \Circle & weak assumption & $\mathcal{O}(n\cdot d)$ \\
\hline
Krum ~\cite{blanchard2017machine} & $n >= 2f + 1$ & \Circle & strong assumption & $\mathcal{O}(n^2\cdot d)$ \\
\hline
multi-Krum \cite{damaskinos2019aggregathor} & $n >= 2f + 3$ & \Circle & strong assumption & $\mathcal{O}(n^2\cdot d)$ \\
\hline
Bulyan~\cite{mhamdi2018hidden} & $n >= 4f + 3$ & \Circle & strong assumption & $\mathcal{O}(n^2\cdot d)$ \\
\hline
Kardam~\cite{damaskinos2018asynchronous} & $n >= 3f+1$ & \CIRCLE & strong assumption & $\mathcal{O}(d+n\cdot f)$ \\
\hline
Zeno~\cite{xie2019zeno} & unbounded & \Circle & weak assumption & $\mathcal{O}(n\cdot d)$ \\
\hline
Zeno+ and Zeno++~\cite{xie2020zeno++} & unbounded & \CIRCLE & weak assumption& $\mathcal{O}(n\cdot d)$ \\
\hline
Draco~\cite{chen2018draco} & $n >= 2f+1$ & \Circle & strong assumption & $\mathcal{O}(n\cdot d)$, run twice \\
\hline
Hogwild! \cite{niu2011hogwild} & unbounded & \CIRCLE & strong assumption & $\mathcal{O}(n\cdot d)$ \\
\hline
ByzSGD~\cite{el2020genuinely} & \multicolumn{1}{c|}{\begin{tabular}[c]{@{}c@{}}$n>= 2f+1$\\ $n >= 3f+2$\end{tabular}} & \CIRCLE & strong assumption & $\mathcal{O}(n^2\cdot d)$ \\
\hline
\textbf{ParSGD} & $n >= 2f+1$ & \LEFTcircle & weak assumption & $\mathcal{O}(n\cdot d)$ \\
\hline
\end{tabular}
\caption{Comparison of Byzantine-resilient GARs. \Circle \ synchronous \CIRCLE \ asynchronous \LEFTcircle \ partial synchronous.}
\label{tab:IntroSum}
\end{table*}

\section{Introduction}
Adversarial attacks against AI and ML attempt to disrupt the training and retraining process by either injecting inaccurate misrepresentative data samples or altering the models’ parameters. In distributed machine learning, each participant trains the global model using its own local data and shares the model’s updates only with central servers or one server. An adversary may compromise workers during the training phase with malicious data samples such as false labels or input. For example, attackers can compromise the mobile devices by sending the junk gradients and parameter updates to the server which severely impact the predicted outcome of the model.  

Distributed Machine Learning (ML) or Federated Learning (FL) \cite{mcmahan2017communication,kairouz2019advances} provides solutions for scaling ML models when training with real-world big data. Usually, with traditional single-data center or distributed data across a range of geographic places, multiple computation parties include the cluster of CPUs, multiple GPUs, or TPUs are aggregated for speed-up and collaboratively train a global model. For advanced distributed ML, the frameworks like Parameter Servers~\cite{li2013parameter}, MapReduce~\cite{dean2008mapreduce}, Graph-parallel~\cite{kalavri2017high}, and All-reduce~\cite{allreduce} are widely used in popular open-source machine learning library including Tensorflow \cite{abadi2016tensorflow}, Horovod \cite{sergeev2018horovod}, Pytorch \cite{paszke2017automatic}, Spark\cite{zaharia2010spark}, etc. 

In distributed training, synchronous and asynchronous modes are used in the above frameworks for data parallelism. Most approaches use stochastic gradient descent (SGD) based algorithms~\cite{bottou2010large}, which are the iterative algorithms for optimizing loss functions to train ML models. For the synchronous method, all workers train the same target model using the different portions of the dataset then take the average of gradients at each iteration (epoch) via All-reduce communication. In asynchronous mode, all workers train the model using their local batch of data independently, then send the updates to a parameter server asynchronously. Several security concerns arise in such a distributed environment since malicious attacks or failures are involved.   \\

Previous works have been studied to mitigate the adversarial attacks in collaborative and distributed learning, \cite{yin2018byzantine},\cite{blanchard2017machine},\cite{mhamdi2018hidden},\cite{damaskinos2018asynchronous},\cite{xie2019zeno},\cite{xie2020zeno++},\cite{chen2018draco},\cite{niu2011hogwild}, and \cite{el2020genuinely}. As we said, in distributed training settings, workers will send updates (gradients) to the server, the server will aggregate all the gradients, calculate the updates, and then return the updated value to all the workers for the next round of training. Workers can be compromised by malicious adversaries or are vulnerable to failures such as software, hardware, and network faults (Byzantine failures). What's more, the averaging aggregation rule like Federated Averaging (FedAvg)~\cite{mcmahan2017communication}, can not tolerate even one outlier because a Byzantine worker can simply tamper its gradient with a wrong direction, leading the averaging vector far away from the correct direction. Current approaches like Krum \cite{blanchard2017machine} are easy to tolerate outliers if the bad gradients are far away from the correct ones. But usually, the malicious workers may pretend they are correct nodes among the system. Some algorithms are not efficient to tolerate the bad gradients that are hidden among the correct ones.  To sum up, building a privacy-preserving and Byzantine-resilient distributed machine learning system becomes extremely important and challenging.\\

In this paper, we analyze and address three main questions: \textit{(i), are there  any reasonable solutions to tolerate at most half Byzantine workers in a proper and explainable way? (ii), if some of the workers are to be crashed, how can we detect and extract them from the total gradients so that the ML models can converge resiliently during the training? (iii), some Byzantine-resilient gradient aggregation rules (GARs) have been proposed and tested under the strong or weak byzantine assumption, but we found some of them may reach good accuracy in strong assumption and bad in weak assumption, and vise versa. Can we create a new aggregation rule which works in both strong and weak assumptions?}   \\

We propose a new GAR solution which is based on the mean of $f$ nearest neighbors to the median among all gradients submitted by workers. We will show that this solution can tolerate at most $f \simeq \frac{n}{2}-1$ Byzantine workers and degrade the time complexity from $\mathcal{O}(d\cdot n^2)$ to $\mathcal{O}(d\cdot n)$ in comparison with Krum, Multi-Krum and so on. Our GAR, ParSGD, can find $g$ as a new median when $f \leq \frac{n-1}{2}$ workers are attacked. After finding this new median, we try to find $f$ gradients closest to this new median and return the mean of $f+1$ gradients (including the median)  as a new parameter for the next epoch training. We define both strong Byzantine resilience and weak Byzantine resilience based on the ParSGD and the formal definitions are listed in Section \ref{sec:proto}. In short, \textit{if $f$ nodes are Byzantine workers and their gradients are far away from the median, we define it as \underline{strong Byzantine resilience} because the aggregation rule will never choose the Byzantine gradients. However, if some Byzantine nodes are pretending to be correct workers and mix up with correct ones, we define it as \underline{weak Byzantine resilience}.} Accuracy will be used as a performance metric for evaluating our ParSGD in comparison with other algorithms. We also propose to use an unknown bounded time $\delta t$ for excluding the crash workers compare the results to other solutions.\\
\\
\textbf{Contribution.} In summary, the three primary contributions of this work are as follows: 

\begin{itemize}[leftmargin=*]
\item We propose a new aggregation rule, named ParSGD, to tolerate Byzantine failures in distributed ML systems. Our experimental results show that the accuracy of ParSGD can reach near $f \simeq \frac{n}{2}-1$ Byzantine workers among $n$ workers with $\mathcal{O}(n\cdot d)$ time complexity. Compared with Mean, Median, and Krum, we get the best accuracy under three common attacks with two different datasets.
\item  Theoretically, we redefine strong and weak Byzantine resilience definitions based on ParSGD, and proved ParSGD can reach a relatively stable accuracy under both of these two assumptions.
\item Our ParSGD is a new partial synchronous GAR based on an unknown bounded time $\delta t$ which can efficiently exclude crash workers to make the training converge faster. Both theoretical and experimental analyses are also provided.
\end{itemize}
~\\
\textbf{Paper Organization.} Section \ref{sec:relted} discusses related work. Section \ref{sec:background} introduces the motivations and some background definitions. Section \ref{sec:proto} introduces our new proposed aggregation rule ParSGD, definitions of strong and weak Byzantine resilience in partial synchronous SGD, time complexity analysis, and proof of its Byzantine resilience. Section \ref{sec:analysis} analyzes the convergence of a distributed SGD using ParSGD. Section \ref{sec:eval} presents our experimental evaluation of ParSGD and some discussion. We summarize our conclusion in Section \ref{sec:conclusion}.

\section{Related Work} \label{sec:relted}

\subsection{Byzantine fault tolerance.}
In previous studies on Byzantine fault tolerate (BFT), several consensus protocols and systems have been proposed to tolerate arbitrary faults in distributed system (e.g., PBFT \cite{castro2002practical}, Raft \cite{ongaro2014search}, BFT-Smart \cite{bessani2014state}, Honey Badger \cite{miller2016honey}, Chios \cite{duan2020intrusion}, etc). All of them need to make a consensus before delivery; however, it is costly in terms of communication overhead for employing consensus in distributed ML architecture. Because it may have thousands of workers processing very large datasets and all workers need to reach a consensus. On the other hand, the design of ParSGD is inspired by the concept of BFT to tolerate Byzantine workers which may submit malicious updates. However, we do not directly adopt BFT since we achieve a slightly different goal: BFT achieves the total order of events while we seek to tolerate malicious updates from the workers. 

\subsection{Byzantine-resilient SGD.}
To tolerate the outliers, robust statistics have been proposed. We summarize them in Table I. Yin et al. \cite{yin2018byzantine} proposed Median \footnote{In the paper, we define the uppercase Median as a GAR solution and the lowercase median as the middle value.} and Trimmed-mean solutions, in which the server sorts all of the gradients and takes median as the global parameter for next round training, same as the Trimmed-mean which needs to remove a percentage of outliers after sorting the gradients. However, a recent paper proved that the Median aggregation rule is still under an order-optimal error rate \cite{fang2020local}. Blanchard et al. \cite{blanchard2017machine} proposed Krum for selecting a valid vector update. This rule has local time complexity $\mathcal{O}(d\cdot n^2)$ which makes it relatively expensive to compute when the $d$ and $n$ are large. If Krum can fully tolerate Byzantine workers, time can be a good trade-off; however, the Krum assumes only one neighbor pretends as a valid vector among the correct vectors, and in each iteration, the algorithm chooses $n-f-2$ neighbors which are not reasonable because Byzantine workers may corrode more workers to pretend they are correct. What's more, Chen et al.\cite{chen2018draco} proposed Draco which leverages a gradient-coding based algorithm for robust learning. EI Mahdi et al. \cite{mhamdi2018hidden} proposed a strong Byzantine algorithm, named Bulyan, which needs $n>=4f+3$ for tolerating fewer Byzantine workers and convergence requires strong assumptions. To preserve training convergence, Sohn et al. \cite{sohn2020election} proposed a voting-based authentication to tolerate inaccurate training results. Alistarh et al. \cite{alistarh2017qsgd} utilize historical information to achieve the best sample complexity of training. Xie et al. \cite{xie2019zeno} proposed Zeno to tolerate the cases where the majority of workers are not fully trusted during training. Xie et al. \cite{xie2018generalized} also demonstrated that Median and Krum exhibit poor performance. Other asynchronous algorithms also have been proposed (e.g., Zeno++ \cite{xie2020zeno++}, Kardam \cite{damaskinos2018asynchronous}, Hogwild! \cite{niu2011hogwild}, ByzSGD~\cite{el2020genuinely}), but inevitably the lack of information on the descent directions will cause a low training accuracy since asynchronous models do not depend on strict arrival times of messages for reliable operations. 

\section{Problem Settings and Background}\label{sec:background}

We consider a large-scale distributed machine learning system with Byzantine-resilient GARs. At the core of our proposed gradient aggregation rule (GAR) ParSGD is to find a new vector $\bm{g}$ as correct medians after $f \leq \frac{n-1}{2}$ workers are attacked in $d$-dimensional vectors. And calculate the mean of $f$ gradients near the median. To better understand our algorithm, in this section, we first compare available approaches of Byzantine-resilient GARs. Then we describe problem settings, relevant attacks, and some background definitions of our ParSGD.

\subsection{Byzantine-resilient GARs.}

Several distributed ML aggregation rules $F(\cdot)$ have been proposed to tolerate Byzantine faults. We summarize some of them in Table \ref{tab:IntroSum}. Median \cite{yin2018byzantine} and Krum \cite{blanchard2017machine} have become the most efficient and effective ones for tolerating Byzantine faults. Authors showed that in theory, both of them can tolerate at most $\frac{1}{2}$ Byzantine workers, we say their \textit{breakdown points} \cite{geyer2006breakdown} can reach $\frac{1}{2}$.  We will show that the sample median is an unbiased estimator which can also reach the \textit{breakdown point} in Section 4. Krum proposed a solution using $n-f-2$ squared-distance-based GAR to find a vector that is closest to the $barycenter$\footnote{https://en.wikipedia.org/wiki/Barycenter} among $n$ vectors.  The algorithm is resilient under random Gaussian attack when the $f$ vectors are far from the correct vectors with a small variance. Fig1(a) shows $f$ Byzantine workers are at a distance far away from the correct workers. As long as the number of correct workers is more than Byzantine workers, then the Krum can always find a correct gradient and sum up its nearby $n-f-2$ neighbors to find a minimum vector among the correct workers. But it can hardly tolerate the Bit-flipping attack which may change all the gradients to opposite directions when there are 50\% Byzantine workers. Fig1(b) shows under Bit-flipping attack where $B$ can never find enough correct neighbors and would to count Byzantine gradients instead, and Krum will return $B$ as the final vector for next round training. Our ParSGD will find the median first, then take the mean of $f$ gradients near the median, as shown in Fig1(c). The Median solution has the same issue, if $n > 2f$, and the variance is large enough, the real median still can be changed to an imprecise gradient even there is only one Byzantine worker, although the new median is among the correct workers. In addition, both Krum and Median are just in search of one optimal vector among $n$ vectors and this vector is vulnerable to be attacked. We will show that these two solutions are not available and have not enough training accuracy under around $\frac{1}{3}$ of Byzantine workers in the evaluation section, even both of their \textit{breakdown points} can reach $\frac{1}{2}$ as shown in Section 6. Furthermore, none of these synchronous GARs can tolerate crash-stop failure. Except one of the GARs, named Hogwild! \cite{niu2011hogwild}, employs a coordinator to monitor the number of workers which is time-consuming. 

\begin{figure}[t] 
\centering 
\includegraphics[width=0.4\textwidth]{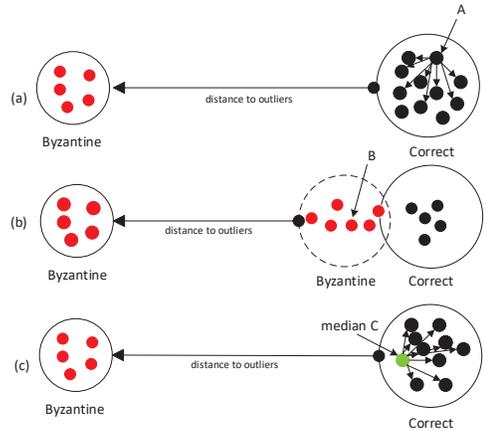} 
\caption{(a) the number of correct gradients larger than Byzantine gradients, (b) some Byzantine gradients pretend they are correct, $B$ becomes the one closest to the $barycenter$ among $n$ gradients, (c) $f$ gradients close to median.} 
\label{Fig.main2} 
\end{figure} 

\subsection{Problem settings}

We consider the problem of minimizing an loss function that has the form of a sum, $Q(w) =\frac{1}{n}\sum_{i=1}^{n} Q_i(w) $. Each of function $Q_i$ is associated with the $i$-th training set. 
The sum-minimization problem also arises for empirical risk minimization (ERM) \cite{zhang2017mixup}. In this case, $Q_{i}(w)$ is the value of the loss function, and $Q(w)$ is the empirical risk. When used to minimize the above function, a standard gradient descent method would perform the following iterations:
\begin{gather}
\bm{w}^{t+1} := \bm{w}^{t} - \gamma \nabla Q(\bm{w}^t) = \bm{w}^t - \frac{\gamma}{n} \sum_{i=1}^{n} \nabla Q_i(\bm{w}^t)  
\end{gather}

where $t$ is the index of epoch, $\gamma$ is learning rate, and the aggregation rule $F=\frac{1}{n} \sum_{i=1}^{n} \nabla Q_i(\bm{w}^t)$.

We assume that $\alpha$ fraction of the $n$ workers are Byzantine and 1 - $\alpha$ fraction are correct workers. The Byzantine workers will not obey the rule of GARs by sending arbitrary messages or pretending they are valid among $n$ correct workers. In addition, correct workers can be crashed down due to software, hardware, or communication issues. In our assumption, we assume that a system can only tolerate at most 50\% of Byzantine workers. If $f$ workers are Byzantine, then $n - f$ correct workers must be larger than $f$, we then get $n > 2f$. When $n \leq 2f$, the GARs still can choose the minority group of gradients and aggregate them for an update, but the result is not under Byzantine resilience assumptions because it breaks the principle of minority versus majority. Under this premise, median is the most appealing unbiased estimator, however, it can be maliciously altered to a new one and no longer be accurate. 

\subsection{Relevant Attacks}
\subsubsection{Crash-stop failure} A crash failure \cite{schneider1984byzantine} occurs when a node suffers from an omission failure once, and then continues to not respond. 
\subsubsection{Bit-flipping Attack} Bit-flipping attack \cite{rakin2019bit}  is an attack which the attacker can change the ciphertext to result in a predictable change  of  the  plaintext. 
\subsubsection{Random Gaussian Attack} Random Gaussian Attack also known as Gaussian Noise \cite{gaussian_noise}, it's a statistical noise having a probability density function equal to normal distribution.

\subsection{Relevant Definitions }
We introduce the coordinate-wise median and some other relevant definitions as follows, which serve as cornerstones for our algorithms. \\
\\
\textbf{Definition 1} \textit{(Coordinate-wise median): For vectors $\bm{V}_i$ in    $\mathbb{R}^d$, $i \in [1, n]$, the coordinate-wise median $\bm{g}=med\{\bm{V}_i : i \in [1, n]\}$ is a vector with its $k$-$th$ coordinate being $g_k = med\{V_i^k : i \in [1, n]\}$ for each $k \in [d]$, where $med$ is a one-dimensional median and $\bm{g}_k$ is one column of k medians. } \\
\textbf{Definition 2} \textit{(Bounded variance) $\forall i \in \{1,..,n\}$, $diag(\mathbb{E}[ (\mathbb{E}\bm{V}_i - \bm{\mu})(\mathbb{E}\bm{V}_i - \bm{\mu})^{T} ] )\leq \sigma^2$. }\\
\textbf{Definition 3} \textit{(Absolute skewness): For one-dimensional random variable $\bm{X}$, define the absolute skewness as $\gamma(X)$ $=$ $\mathbb{E}[(\frac{X - \mu}{\sigma})^3]$ $=$ $\frac{\mathbb{E}[(X - \mu)^3]}{\sigma^3}$ $=$ $\frac{\mathbb{E}[(X - \mu)^3]}{(\mathbb{E}[(X - \mu)^2])^{3/2}}$ $=$ $\frac{K_3}{K_2^{3/2}}$, and let $S_k$ be the skewness of random distribution, say $S_k = \frac{m_3}{m_2^{3/2}} = \frac{\frac{1}{n}\sum_{i=1}^{i=n}{(x_i-\bar{x})^3}}{[\frac{1}{n}\sum_{i=1}^{i=n}{(x_i-\bar{x})^2}]^{3/2}}$, where $\bar{x}$ defined as sample mean, $m_2$ as the second central moment(variance), and $m_3$ defined as third central moment.} \\
\textbf{Definition 4} \textit{(Lipschitz continuity): $\forall(w, w')$, $h$ is $L$-Lipschitz if $\|h(w) - h(w')\| \leq L\|w - w')\|_2$}  \\
\textbf{Definition 5} \textit{(Smoothness): $h$ is $L'$-smooth if and only if $\forall(w, w')$, $\|h'(w) - h'(w')\|_2 \leq L'\|w - w')\|_2$ } \\
\textbf{Definition 6} \textit{(Strong convexity): $h$ is $\mu$-strongly convex if and only if  $\forall (w, w')$, $h(w')$ $\geq$ $h(w) + h'(w)^{\top}(w'- w) +\frac{\mu}{2}\|w'-w\|_2^2.$  }  \\

\section{Partial Synchronous Byzantine-resilient SGD} \label{sec:proto}

We now introduce our partial gradient aggregation rule, ParSGD, which satisfies both weak and strong Byzantine resilience. The detailed distributed synchronous SGD is shown in \textit{\textbf{Algorithm 1}}.

In a normal distribution, the median equals to mean. Namely, if $|g_k - \mu_k|$ is relatively small, then we can achieve good training accuracy. \textit{\textbf{Theorem 1}} proved that the \textit{breakdown point} of median can reach 50\%, but it doesn't mean it will always be 50\% because it can easily be altered if Byzantine workers are pretending. In this section, we proposed two corollaries, \textit{\textbf{Corollary 1\&2}},  which help explain why to take the mean of closest gradients near median is an excellent solution.\\
\\
\textbf{Theorem 1.} \textit{With the fraction of Byzantine workers as $\alpha$ and the sample median as the estimator, the asymptotic breakdown point \cite{geyer2006breakdown} for $\alpha$ is $1/2$.}

\begin{proof} If we have $m$ data points and we let a minority of $\lfloor \frac{ m-1}{2} \rfloor$ points become outliers leaving the rest of the fixed points, and the median stays with the majority. Usually median may change, but it does not become arbitrarily bad, then the sample breakdown point is
\begin{align}
    \frac{\lfloor \frac{ m-1}{2} \rfloor}{m} = \lfloor \frac{m-1}{2m} \rfloor  \notag \\
    \Rightarrow  \frac{1-\frac{1}{m}}{2}, \notag 
\end{align}

the more $m$ we have, the breakdown point will be closer to one-half. The asymptotic breakdown point is one-half.
\end{proof}

\subsection{Important Notations}
The notations used in this paper is summarized in Table 2.
\begin{table}[th]
\footnotesize
\begin{center}
        \begin{tabular}{|c|c|}
        \hline
        \!\!\textbf{Notation}\!\!& \textbf{Meaning}\\
        \hline
        \hline
        $i$ & index of worker \\
        \hline
        $n$ & number of total workers \\
        \hline
        $T$ & number of epochs \\
        \hline
        $t$ & index of epochs \\
        \hline
        $\delta t$ & unknown bounded time \\
        \hline
        c & number of crash workers \\
        \hline
        $f$ & number of Byzantine workers \\
        \hline
        $\gamma$ & learning rate \\
        \hline
        $\bm{V}$  & all vector of d-dimensional gradients \\
        \hline
        $\bm{U}_{f+1}$  & $f$ vectors near median, including the median \\
        \hline
        $\bm{\bar{U}}$  & take the mean of  $\bm{U}_{f+1}$ \\
        \hline
        $g_k$  & the value of median in $k$-th dimension\\
        \hline 
        $\bm{g}$  & a vector of selected median \\
        \hline 
        $\mu_k$  & average value in $k$-th dimension\\
        \hline 
        $F$  &  gradient aggregation function\\
        \hline 
        \end{tabular}
\end{center}
\caption{Important Notations}
\label{tab:notation}
\end{table}

\begin{algorithm}[ht]
\SetAlgoLined
~\\
\textbf{Workers: $i = 1,...,n$}\\
  \For{$t = 1,..., T$}{
  Waiting and receive $\bm{w}^{t}$ from the server\;
  Training, compute, and send the new gradients $\bm{V}_i^{t+1} = \nabla Q_i^t(\bm{w}^{t}) $  to the server
 }
 ~\\
 \textbf{Server:}\\
 input: $\delta t$ (Empirically and manually initialize a enough time $\delta t$) \;
\For{$t = 1,...,T$}{
Broadcast $\bm{w}^{t}$ to all the workers\;
Wait only $2*\delta t$ bounded time for $n-c$ gradients $\{\bm{V}^t_i: i \in [n-c]\}$ arrive\;
Recalculate and update the $\delta t$ if all the gradients are being collected\;
Calculate $f$ as defined in Definition 7, $f = \frac{n-c-1}{2}$\;
Compute $\bm{\bar{U}}$ as defined in Corollary 1 and 2\;
Update the parameter $\bm{w}^{t+1} = \bm{w}^{t} -  \gamma^t \bm{\bar{U}}$\;
}
\caption{Partial Synchronous Byzantine SGD}
\end{algorithm}

\subsection{Partial Synchrony}
We first consider timing assumption in partial synchronous SGD. In a synchronous system, there is a known fixed upper bound $\delta t$\footnote{We define $t$ represents the index of an epoch, and $\delta t$ represents a bounded time.} on the time required for a message to be sent from one worker to another. In an asynchronous system, no fixed upper bound $\delta t$ exists. And partial synchrony \cite{dwork1988consensus}, fixed bound $\delta t$ exists, but it's not known a priori\footnote{https://en.wikipedia.org/wiki/A\_priori\_and\_a\_posteriori}.\\
\\
\textbf{Proposition 1.} \textit{If a trusted GAR executes aggregation $F(\cdot)$ during a bounded and not-a-priori time $\delta t$, then $n-c$ active vectors are eventually collected, where $n$ refers to the number of workers, and $c$ refers to the crash nodes.}\\

In this proposition, the server only needs to initialize an estimated time $\delta t$ which is enough for collecting all the gradients in one round. In the rest of the epochs, the GAR will recalculate the aggregation time $\delta t$, and only wait for $2*\delta t$ bounded time which is not known a priori. 

\begin{proof} We initially set an unknown bounded time $2 * \delta t$ to collect $n-c$ active workers, the proof is simple and straightforward. We assume $c$ number of workers may be crashed or compromised by malicious failures (crash-stop failure is also a kind of Byzantine failure). In this case, Byzantine workers may spend more training time than normal workers because of the extra time for Byzantine infection. Simply, we only wait for $2*\delta t$ time in aggregation for the crash nodes, eventually, $n-c$ vectors will be collected. We exclude $c$ crash workers which may be caused by network delay, malicious attack, or arbitrary system failures. 
\end{proof}

\subsection{Byzantine Resilience}

We introduced strong and weak Byzantine resilience in Section 1. In order to explain how our ParSGD work both in strong and weak Byzantine resilience, we formally define our strong and weak Byzantine resilience GARs in partial synchronous SGD and two Corollaries as follows.\\

\textbf{Definition 7} \textit{(Weak Byzantine Resilience): Given a system of $n$ components, \(b\) of which are Byzantine workers, including $c$ random crash-stop vectors as $\bm{C}_1$,...,$\bm{C}_c$ and $f$ malicious workers pretending as corrects ones as {$\bm{B}_1,...,\bm{B}_f$}, where $0 \leq b \leq n, b= f+c,$ and $ b,c,f \in \mathbb{N}$. Let $\bm{V}_1$,...,$\bm{V}_{n-b}$ be the  independent identically distributed(i.i.d.) random in $\mathbb{R}^d$, $\bm{V}_i$ $\sim \mathcal{N}(\bm{\mu},\,\bm{\Sigma})$, and with $ \mathbb{E}V_i^k=\mu_k $, $k$ refers to dimensional index. $F$ is said to be weak Byzantine resilience if it satisfies} 

\centerline{$F= F\underbrace{(\overbrace{\bm{V}_1, ..., \underbrace{\bm{B}_1,...,\bm{B}_f,}_{f}..., \bm{V}_{n-f-c}}^{n-c}, \underbrace{\bm{C}_1,...,\bm{C}_c}_{c})}_{n}$
} 
~\\
(i) $n > 2f + c$, and \\
$\begin{cases}
f = 0, \ c = 0 & \quad \text{if \ Byzantine free}\\
0 \leq c < n,\ 0 \leq f \leq \lceil \frac{n-c-1}{2} \rceil  & \quad \text{if } \text{$f+c > 0$}
\end{cases}$ \\
(ii) $|g_k - \mu_k| \leq \epsilon$,\textit{ where $\epsilon$ is negligible, $g_k$ refers to the median.} \\
\\
\textbf{Definition 8} \textit{(Strong Byzantine Resilience):  $F$ is said to be strong Byzantine resilience if it satisfies condition (i) and (ii) in Definition 7, and  
(iii) $|g_k - V^{far}_i| < |g_k - B^{close}_j|$, where $g_k$ is the median, $V^{far}_i$ refers to the farthest correct gradient, and $B^{close}_j$ refers to the closest Byzantine gradient.}  \\

Theorem 1 proved Median is an unbiased estimator which can tolerate at most $\frac{1}{2}$ Byzantine workers. Even there are $f \simeq \frac{n}{2}-1$ Byzantine workers, Median can still find a new correct gradient as the new median to tolerate the arbitrary gradients attack. Based on Theorem 1 and Definition 7 and 8, we get two corollaries below.\\
\\
\textbf{Corollary 1}: \textit{Under the assumption of Definition 7, 8 and Theorem 1, suppose we have random vectors $\bm{\tilde{V}}_i \in \mathbb{R}^d, i \in [1, n-c]$ containing normal gradients and malicious ones. The set of $f$-neighbor vectors of $\bm{g}$ is defined as $\bm{U}_{f+1} := \{ \{\bm{\hat{V}}_i\}_{i=1}^f, \bm{g} \}$, where $\{\bm{\hat{V}}_i\}_{i=1}^f = \arg \min \sum_{i=1}^f \sum_{k=1}^d |V_{(i)}^k -g_k|$, for different $V_{(i)}^k \in V_j^k, i\in[1, f], j\in[1, n-f-c] $. Here $\bm{U}_{f+1}$ is obtained by comparing the distance between each vector and median per dimension. We can conclude that none of the vectors in $\bm{U}_{f+1}$ contains Byzantine values. The estimator to update the gradient is $\bm{\bar{U}} := \frac{1}{f+1}(\sum_{i=1}^{f+1} \bm{\hat{V}}_i + \bm{g})$}. 
\begin{proof} Suppose we have a set $S\!:=\!\{\bm{V}_1,...,\bm{V}_{n-f-c},\bm{B}_1, ...,\bm{B}_f\}$ taking value in  $\mathbb{R}^d$. In the $k$-th  dimension, let $V_1^k,...,V_{n-f-c}^k$ and $B_1^k,...,B_f^k$ be the reordered gradients such that $|V_1^k - g_k| \leq ... \leq |V_{n-c}^k - g_k|$ and $|B_1^k - g_k| \leq ... \leq |B_f^k - g_k|$ , where $g_k$ refers to the sample median. Under the assumption of Definition 7 and 8, $|g_k - V^{far}_i| < |g_k - B^{close}_j|$, meaning that 
\begin{gather}
|V_1^k\!-\!g_k| \leq...\leq |V_{n-f-c}^{k}\!-\!g_k| \leq |B^k_1\!-\!g_k| \leq ... \leq |B_f^k\!-\!g_k|  
\end{gather}
Definition 7(i) limits the $f$ must smaller than $n-f-c$. We defined 
\begin{gather}
     n > 2f + c \Rightarrow n >= 2f + c + 1 \notag \\
    then, f <= \lceil \frac{n - c - 1}{2} \notag \rceil
\end{gather}
We take the largest $f$ in $n$ workers,
\begin{gather}
    n = 2f + c + 1 \notag \\
    n - f - c = f + 1  \notag
\end{gather}
Here, $f+1$ equals to $n-f-c$,  we get the first $n - f -c$ \textbf{correct} gradients and append them to the $\bm{U}_{f+1}$.
\end{proof}
~\\
\textbf{Corollary 2}: \textit{Under the assumption of Definition 7 and Theorem 1, suppose we have random vectors $\bm{\tilde{V}}_i \in \mathbb{R}^d, i \in [1, n-c]$ containing normal gradients and malicious ones. The set of $f$-neighbor vectors of $\bm{g}$ is defined as $\bm{U}_{f+1} := \{ \{\bm{\hat{V}}_i\}_{i=1}^f, \bm{g} \}$, where $\{\bm{\hat{V}}_i\}_{i=1}^f = \arg \min \sum_{i=1}^f \sum_{k=1}^d |V_{(i)}^k -g_k|$, for different $V_{(i)}^k \in V_j^k, i\in[1, f], j\in[1, n-f-c] $. We can conclude that $V_{(i)}^k$ may contain Byzantine vector elements, but all the vectors in $\bm{U}_{f+1}$ still move in the bounded deviation direction to make the training converge. The estimator to update the gradient is $\bm{\bar{U}} := \frac{1}{f+1}(\sum_{i=1}^{f+1} \bm{\hat{V}}_i + \bm{g})$}  

\begin{proof} Since some Byzantine workers pretend to be correct ones, the median vector $\bm{g}$ is important for the gradient direction. As we know the breakdown point of $\alpha$ is $\frac{1}{2}$, we will show the $\mathbb{E}[\bm{g}] = \bm{\mu}$. Because we perform the median selection per coordinate, the proof only needs to show $\mathbb{E}[g_k] = \mu_k$, on the $k$-th dimension, with $k \in [1, d], k \in \mathbb{N}$.

Considering a set of i.i.d normal variable $\{X_j\}_{j=1}^m$ and $X_j \in \mathbb{R}$, $\mu$ and $\sigma^2$ are the mean and variance respectively. Let $X_{j:n}$ be the $j$-th order in the a sorted sequence $\{X_{1:m}, ..., X_{m:m}\} = \{X_j\}_{j=1}^m$. Suppose $m$ is odd, the median of the set is $M := X_{\frac{m-1}{2}:m}$, which has the probability density of this order statistics as:

\begin{align}
    f_{med}^{odd}(x) = \frac{m+1}{2} {m \choose \frac{m-1}{2}}f(x)(F(x)(1-F(x)))^{\frac{m-1}{2}}
\end{align}

where $f(x)$ and $F(x)$ are the density and cumulative distribution functions respectively.
Since the normal distribution is symmetric, $F(x)=1-F(2\mu -x)$, we can easily get $f_{med}^{odd}(x) = f_{med}^{odd}(2\mu - x)$, namely
\begin{align}
    \mathbb{E}[X_{\frac{m-1}{2}:m}] = \mu \label{odd}
\end{align}

When $m$ is a even number, the median is $M := \frac{1}{2}(X_{\frac{m-1}{2}:m} + X_{\frac{m-1}{2}:m})$. So the joint probabily density function is:
\begin{align}
    f_{med}^{even}(x_1, x_2) = (\frac{m}{2})^2{m \choose \frac{m}{2}}f(x_1)f(x_2)(F(x_1)(1-F(x_2))^{m-1}
\end{align}

Again, using the symmetric property, we have
\begin{align}
    \mathbb{E}[M] = \mathbb{E}[2\mu - M] \label{even}
\end{align}

which means $\mathbb{E}[M] = \mu$, still. From eq.(\ref{odd}) (\ref{even}), we know that the sample median of a normal distribution is an unbiased estimator. When Byzantine samples existing, provided that $\alpha<\frac{1}{2}$, we can easily see the sample median as the unbiased estimator still holds.

We then prove the weak Byzantine resilience. Under the assumption of Definition 7(ii) $|g_k - \mu_k| \leq \epsilon$, where $\epsilon$ is negligible. The $\bm{V}_i$ in a probability distribution is symmetric about the mean, presuming that data near the mean are more frequent in occurrence than data far from the mean. Even some Byzantine workers pretend they are correct among $V_i^k$, they still have high likelihood values with a given normal distribution, and we say the sample mean of the set $\bm{U}_{f+1}$ still in the right direction. 
\end{proof}

\subsection{Complexity Analysis}
Computing the median of a list of $n$ of unordered elements takes $\mathcal{O}(n)$ time with standard selection algorithms for a $1$-dimensional vector, and the time complexity to get the median which is only $\mathcal{O}(1)$. ParSGD has one more step than Median, which is to find $f$ closest gradients, and the time complexity of this step is $\mathcal{O}(n)$. Usually, there are $d$-dimensional vectors, so the overall time complexity is $\mathcal{O}(d\cdot n).$

\section{Convergence Analysis} \label{sec:analysis}
In this section, we analyze the convergence of the SGD using ParSGD algorithm defined in Section 4. All vectors in $\bm{U}_{f+1}$ are in the normal distribution, and all the elements of the vectors are from correct values in strong Byzantine assumption. The absolute value between $k$-th dimension of $\bm{\bar{U}}$ and $\mu_k$ is bounded in a small value ($\eta$) which will help lead a right direction of gradient descent. Same as the weak Byzantine assumption, some workers may pretend they are correct gradients in $\bm{U}_{f+1}$, but the $f+1$ gradients in $\bm{\tilde{V}}_i$ are pre-selected through $f$ closest neighbors near the median, and we proved median is an unbiased estimator which can tolerate at most $\frac{1}{2}$ Byzantine workers. In Proposition 2, we bounded $\eta$ and found it is only related to the maximum of covariance $\sigma$ and $f$.\\
\\
\textbf{Proposition 2.} \textit{Suppose we have random vectors $\bm{\tilde{V}}_i \in \mathbb{R}^d, i \in [1, n-c]$ containing normal gradients and malicious ones. We assume that $\bm{V}_i$ be any independent and i.i.d $d$-dimensional vector, $\bm{V}_i$  $\sim \mathcal{N}(\bm{\mu},\bm{\Sigma})$, and $g_k$ refers to the median of $\tilde{V}_i^k$. $\sigma$ is the upper bound of the variance under Definition 2. If $n - c > 2f$ and $|g_k - \mu_k| \leq \epsilon$, let $u_k = \bm{\bar{U}}^k$ be the $k$-th dimensional sample mean of $\bm{U}_{f+1}$. We defined $\eta$ by} 
\begin{gather}
\mathbb{E}(u_k - \mu_k)^2 = \frac{f\sigma^2}{(f+1)^2}, \quad \eta = \frac{\sqrt{f}\sigma}{f+1}\leq \frac{\sigma}{2} \notag 
\end{gather}

\begin{proof} Under strong Byzantine resilience, $\bm{\tilde{V}}_i$ only composes of $\bm{V}_i$. All the computation is under a normal distribution condition. The bounded variance can be written as
\begin{gather}
\mathbb{E}(V_i^k-\mu_k)^2 \leq \sigma^2   \\
\mathbb{E}((V_i^k)^2 + \mu_k^2 - 2V_i^k\mu_k ) \leq \sigma^2 \Leftrightarrow \mathbb{E}((V_i^k)^2) \leq \sigma^2 + \mu_k^2 \label{EV_i^k}
\end{gather}

Also, $u_k = \bm{\bar{U}}^k$ and $\mathbb{E}(u_k - \mu_k)^2$ can be calculated as
\begin{align}
     u_k&=\frac{1}{f+1}(\sum_{i=1}^{f+1} V_i^k + g_k) \notag \\
       &= \frac{1}{f+1}(\Delta + g_k),  \notag \\
    where, \Delta &= \sum_{i=1}^{f+1} V_i^k 
\end{align}
\begin{align}
 then, \mu^2_k\!+\!u_k^2\!-\!2u_k\mu_k\!=\!\frac{1}{(f+1)^2}(\Delta^2 + g_k^2 \notag \\
 + 2\Delta g_k)\!+\!\mu^2_k\!+\!\frac{2}{f+1}(\Delta\!+\!g_k)\mu_k  \\
 \mathbb{E}(u_k - \mu_k)^2 \leq \mathbb{E}(\frac{1}{(f+1)^2}(\Delta^2 + g_k^2 + 2\Delta g_k) \notag \\
 + \mu^2_k + \frac{2}{f+1}(\Delta + g_k)\mu_k) \label{Eall}
\end{align}

As defined condition (ii) in Definition 7, $|g_k - \mu_k| < \epsilon$, then
\begin{gather}
\mathbb{E}(g_k\!-\!\mu_k)^2\!=\!\mathbb{E}[g_k^2]\!-\!\mu_k^2 \leq \epsilon^2  \Leftrightarrow \mathbb{E}[g_k^2] \leq \epsilon^2 + \mu_k^2 \label{Egk}\\
\mathbb{E}{(\Delta)}\!=\!f\mu_k,\quad\mathbb{E}(\Delta)^2\!=\!f\cdot(\sigma^2\!+\!\mu_k^2)\!+ \!\frac{f\cdot(f-1)}{2}\mu_k^2 \label{EDelta}
\end{gather}

We finally bounded the $u_k$ to mean combined with formula \ref{EV_i^k}, \ref{Eall}, \ref{Egk} and \ref{EDelta},
\begin{align}
\mathbb{E}(u_k - \mu_k)^2 \leq \frac{1}{(f+1)^2}(f\cdot(\sigma^2 + \mu_k^2) + f\cdot(f-1)\mu_k^2 \notag \\ 
+ \mu_k^2 + \epsilon^2 + 2f\mu_k^2) - \mu_k^2 \notag \\
\leq \frac{f\sigma^2 + \epsilon^2}{(f+1)^2} \Leftrightarrow \frac{f\sigma^2}{(f+1)^2} \notag
\end{align}

\begin{figure*}[ht] 
\centering 
\includegraphics[width=1\textwidth]{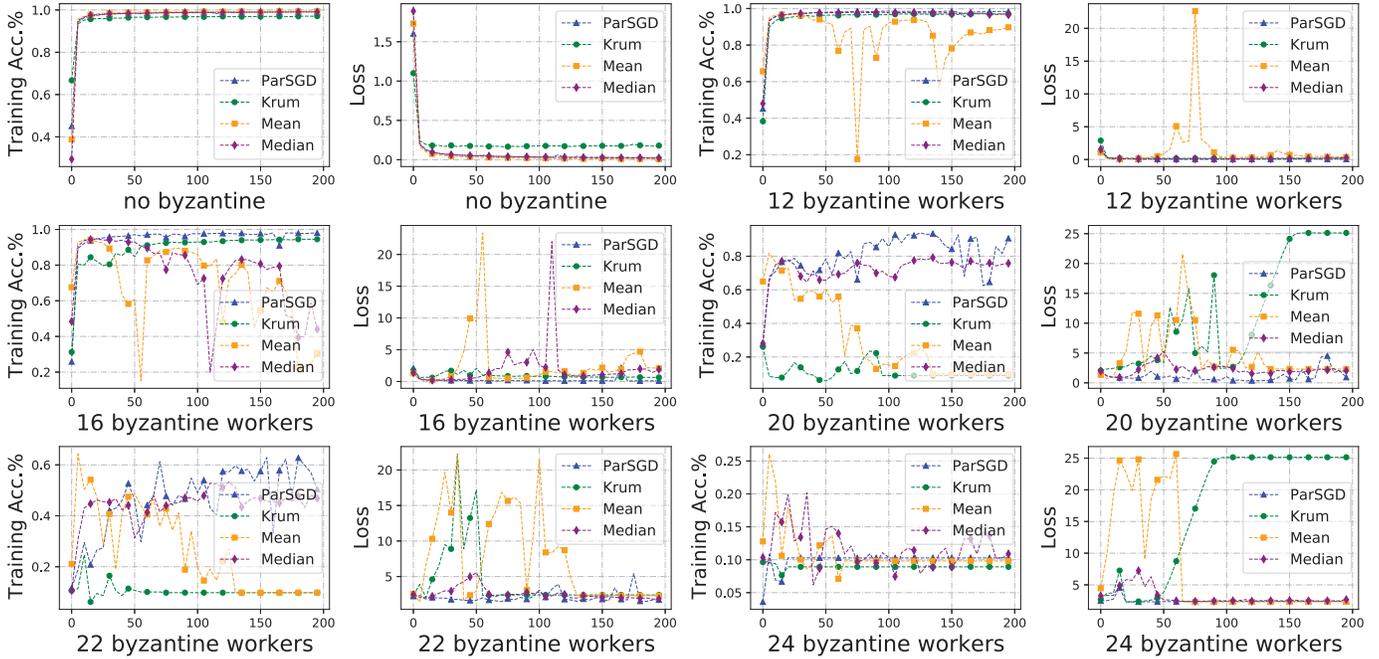} 
\caption{MNIST: Top-1 Accuracy and Loss under Bit-Flip Attack} 
\label{Fig.main2} 
\end{figure*}

\begin{figure*}[ht] 
\centering 
\includegraphics[width=1\textwidth]{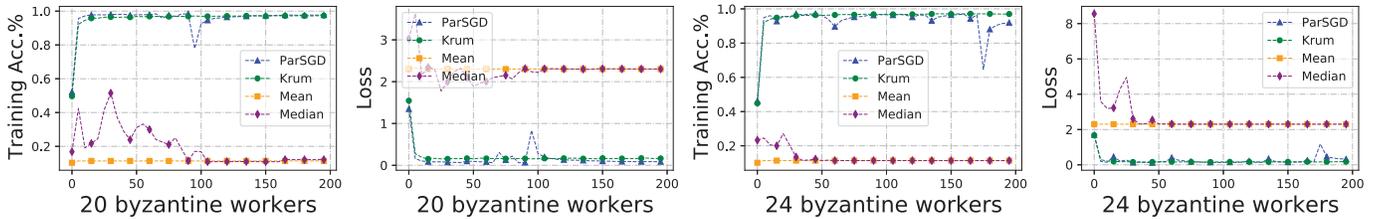} 
\caption{MNIST: Top-1 Accuracy and Loss under Random Gaussian Attack} 
\label{Fig.main3}
\end{figure*}

then standard deviation of $\mathbb{E}(u_k - \mu_k)^2$ equals to  $\frac{\sqrt{f}\sigma}{f+1}$ and $\eta$ is bounded by $\sigma$ and $f$ in no more than $1/2\sigma$ when $f = 1$, and the more workers we have the smaller $\eta$ is, then we say $\eta \leq 1/2\sigma$. 

For the Weak Byzantine resilience, $\tilde{V}_i^k$ may contain $V_i^k$ and $B_i^k$. However, $B_i^k$ pretends to be $V_i^k$, which means $\mathbb{E}(\tilde{V}_i^k-\mu_k)^2 \leq \sigma^2$. Then the $u_k$ is updated  as $u_k = \frac{1}{f+1}(\sum_{i=1}^{f+1} \tilde{V}_i^k + g_k) = \frac{1}{f+1}(\Delta + g_k),$ where$ \ \Delta = \sum_{i=1}^{f+1} \tilde{V}_i^k$. Since $B_i^k$ pretends to be correct, we assume $\mathbb{E}B_i^k = \mu_k$. $\mathbb{E}{(\Delta)}$ and $ \mathbb{E}(\Delta)^2$ remain the same, which means the overall result remains the same. Now that the proposed gradient estimator $\bm{\bar{U}}$ is bounded in a small deviation from the ground truth, the proposed algorithm will converge as the regular gradient descent with mean as the estimator. So we have proved ParSGD algorithm is Byzantine resilience and the training will converge even it is a weak Byzantine resilience.
\end{proof}

\begin{figure*}[ht] 
\centering 
\includegraphics[width=0.9\textwidth]{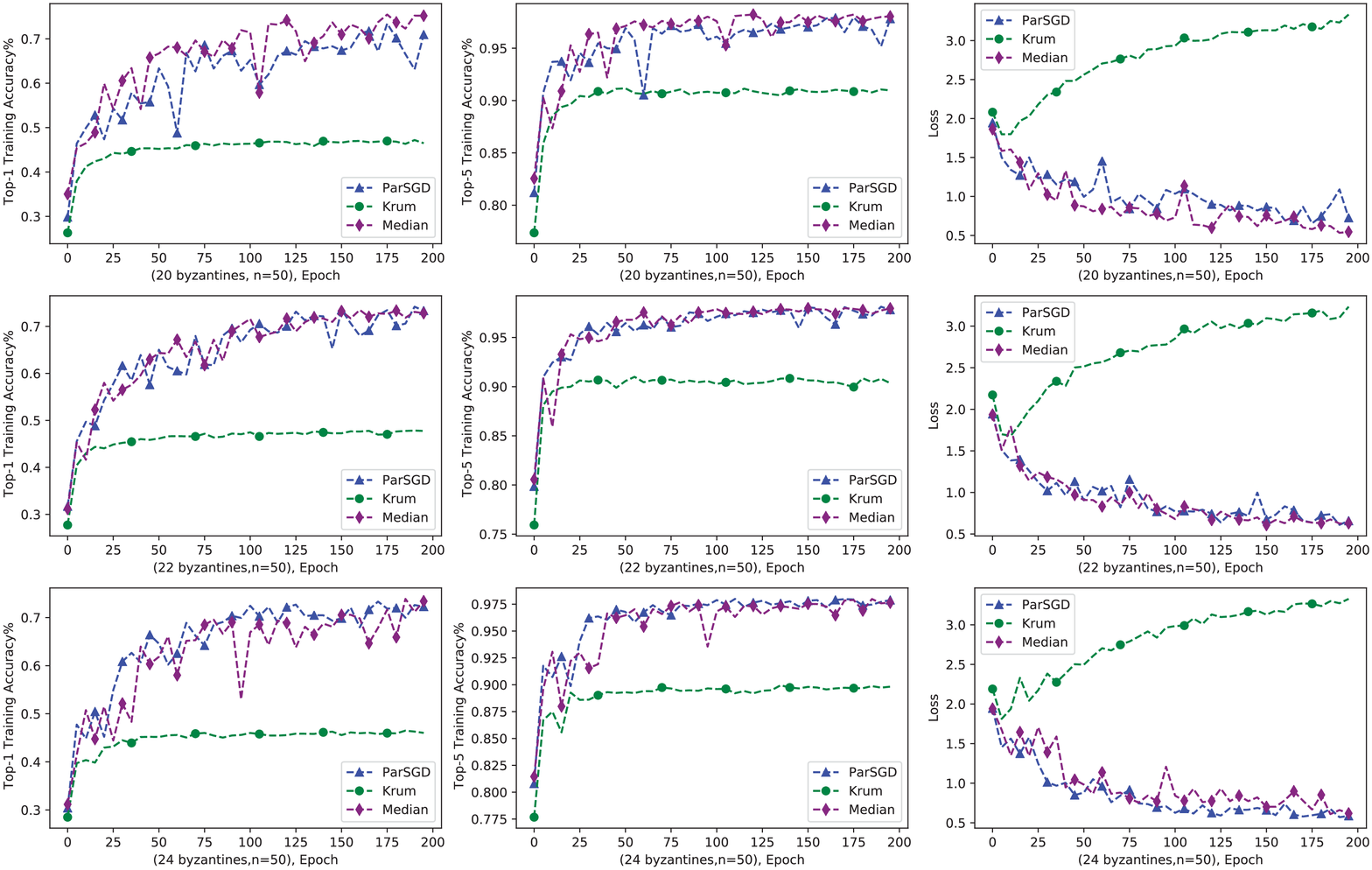} 
\caption{CIFAR10: Top-1, Top-5 Accuracy and Loss under Random Gaussian Attack.}
\label{Fig.main4}
\end{figure*}

\section{Experimental Evaluation}\label{sec:eval}
We implemented and evaluated ParSGD in a simulated mode on a Tesla P100 Nvidia setting. Our algorithm is evaluated by 3 common attacks, Crash-stop  \cite{schneider1984byzantine}, Bit-flipping \cite{rakin2019bit} and Random Gaussian attack. We summarize the results of our experiment as follows:

\begin{itemize}[leftmargin=*]
\item ParSGD does not wait for all workers' updates because it collects gradients in partial synchronous mode. It reaches the best training accuracy under the Crash-stop failure compared with Mean, Median, and Krum. 
\item Under the attack of Bit-flipping, ParSGD gets the best Top-1 accuracy among Mean, Median, and Krum in MINST. When the Byzantine workers $f$ reach 22 (44\% Byzantine workers), some fluctuations occur in Top-1 accuracy affected by the median, but the Loss still converges best. Krum and Mean cannot converge both in CIFAR10 and MINST when Byzantine workers $f$ reach 20, as shown in Figure 2 and Figure 4.  

\item With Random Gaussian attack, compared with Krum which has the inherent advantage of resistance to random Gaussian attack when Byzantine workers $f < \frac{n}{2}$, ParSGD still achieves the best training accuracy and Loss benchmarking with both MNIST and CIFAR10.
\end{itemize}

\subsection{Overviews}

\subsubsection{Datasets}
We conduct experiments on benchmark MNIST \cite{lecun1998gradient}, handwritten digits for image processing, which contains 60,000 training images and 10,000 testing images. We also conduct experiments on benchmark CIFAR10 \cite{krizhevsky2009learning}, which consists of 60,000 32x32 colour images in 10 classes, with 6,000 images per class.
\subsubsection{Evaluation Settings}
During the evaluation, we execute 200 epochs with 50 workers. We set the learning rate to 0.05 and the batch size to 100. In each epoch, we use cross-entropy loss function with Top-1 and Top-5 accuracy evaluation metrics. For the distributed ML model, we run a multi-layer convolutional network, which has four 3x3 convolution layers (the first two layers with 64 channels, the last two layers with 128, each followed with a 2x2 max pooling), and a fully connected layer with 128 units and ReLu activation, with a last output layer.

We normalize all data and divide the data into two classes: training data (eighty of all data), and testing data (twenty of all data).

\subsection{Crash-stop failure}
Crash-stop failure usually happens at a sudden stop in an emergent situation, and it is a type of failure that causes the component of a system to stop operating. We tested ParSGD under 5 crashed workers and 22 Byzantine workers. The result shows that only ParSGD can tolerate Crash-stop failure among Median, Mean and Krum, because ParSGD is in partial synchronous mode which only collects $n-c$ vectors from active workers. 

\subsection{Bit-flipping Attack}
The formal definition of a Bit-flipping attack is an attack in which the attacker can change the ciphertext to result in a predictable change of the plaintext. In our experiment, we simulate the Bit-flipping attack by changing $f$ vectors to opposite descent directions. More specifically, the adversary $p_i$ first calculates each true gradient vector $\Delta w_j$, and then updates $-c_i\Delta w_j$ to the server, where $c_i$ is a random constant (or one for simple). The result is shown in Figure \ref{Fig.main2}.

We evaluate the number of Byzantine workers $f$ from 12 to 24. We found Mean cannot tolerate even one Byzantine worker, while Median and Krum cannot converge at nearly 18 Byzantine workers. Our ParSGD can still converge when Byzantine number $f$ reaches up to 22, see the training accuracy of 22 Byzantines in Figure \ref{Fig.main2}. The accuracy of ParSGD is influenced by the median (median may pretend correctly), but the Loss still converges best even Byzantine number $f$ reaches up to 24, see the Loss of 24 Byzantines in Figure \ref{Fig.main2}.

\subsection{Random Gaussian Attack}
We simulate using random Gaussian attack to compromise the above 4 aggregation rules. We take the standard deviation to 0.1, 1, and 200, and take the mean to $-1e8$ or 0. Our experimental results prove that as long as \textit{the majority of the workers are correct}, the accuracy of ParSGD can always be stable. We found Krum is the best solution here because of its inherent advantage of finding the minimum vector among $n-f$ vectors in strong Byzantine resilience (Definition 8), as shown in Figure \ref{Fig.main3}. However, in weak Byzantine resilience (Definition 7), if all the Byzantine workers pretend they are correct, the training accuracy of Krum will be increasingly threatened. ParSGD still works well when there are 24 Byzantine workers and both of their Top-5 accuracies can reach up to 99\%. Due to space limitations, we only show the Top-1 results here since the Top-5 results have roughly the same conclusion.

We also benchmark ParSGD with CIFAR10 under the random Gaussian attack, as shown in Figure \ref{Fig.main4}. We only list ParSGD, Krum, and Median here, as we have proved that Mean can not tolerate even one Byzantine worker. In the meantime, all the algorithms can tolerate less 30\% Byzantine workers except Mean, so we only show our results for 20, 22, and 24 Byzantine workers here ($n = 50$ and $c = 0$). We evaluate both the Top-1 and Top-5 accuracy in this CIFAR10 evaluation.
 
We compromised $f$ $d$-dimensional vectors by replacing $f$ wrong vectors. These vectors are in normal distribution by setting mean to 0 and variance to 200, pretending all the generated gradients are correct. The results show that the accuracy of Median and ParSGD are almost the same in Top-1 and Top-5, but ParSGD has relatively fewer fluctuations when the Byzantine number reaches up to 24. In addition, ParSGD converges best as shown in Figure 4, see the Loss of 24 Byzantines.

In the nutshell, consider all evaluations, we found only Median and ParSGD get converged correctly. As we said, Krum can hardly converge even Byzantine number $f$ less than correct workers in weak Byzantine resilience.

\subsection{Discussion and Limitation}
During the evaluation, we find that the training accuracy of ParSGD is easily affected by the median because our algorithm is finding $f$ nearest neighbors close to the median. When there are more than 45\% Byzantine workers under Bit-flipping attack and random Gaussian, the median can easily be altered to a wrong one. Namely, if the median is pretty close to the mean in an unknown distribution, we can achieve good training accuracy. But in real-world training settings, the vectors are not completely in a normal distribution, that's why the accuracy starts to fluctuate when Byzantine workers near 50\%. For example, under the Gaussian attack when the variance is relatively normal, $f$ Byzantine workers will pretend they are correct, so a wrong gradient of a Byzantine node may become be selected to be a new median. The result of collecting $f$ gradients near median may count a lot of Byzantine workers, but because they pretend to be normal, the training finally gets converge, however, it just has some fluctuations during training. Compared with Krum and Median, our algorithm ParSGD performs a good result when there are more than $\frac{1}{3}$ Byzantine workers. We demonstrate our algorithm, ParSGD, is practical and meaningful in an untrusted distributed environment.

\section{Conclusion}\label{sec:conclusion}
We propose a novel gradient aggregation rule, ParSGD, in partial synchronous mode, which can tolerate Crash-stop failures and nearly 50\% of Byzantine workers in distributed and collaborative training. Compared with Mean, Median, and Krum, we get the best accuracy under three common attacks. The algorithm has a provable convergence analysis in both strong and weak assumptions. We will try to apply this solution to federated learning settings in the future.

\section*{Acknowledgements}
We gratefully acknowledge the support of the NSF through grant IIP-1919159. We also acknowledge the support of the IBM research team.   \\

\bibliographystyle{abbrv}
\bibliography{ref}

\end{document}